\documentclass{iau} 
\usepackage{graphicx}

\title[~~Multiple stellar populations in star clusters] 
{Multiple stellar populations: from old Milky Way globulars to young star clusters}

\author[Anna F. Marino]   
{Anna F. Marino$^{1,2}$}
\affiliation{$^1$Dipartimento di Fisica e Astronomia ``Galileo Galilei'' \\ Univ. di Padova, Vicolo dell'Osservatorio 3, Padova, IT-35122 \\ email: {\tt anna.marino@unipd.it} \\[\affilskip]
$^2$Centro di Ateneo di Studi e Attivita' Spaziali ``Giuseppe Colombo'' \\ CISAS, Via Venezia 15, Padova, IT-35131}

\pubyear{2019}
\volume{351}  
\jname{Star Clusters: From the Milky Way to the Early Universe}
\editors{A. Bragaglia, M.B. Davies, A. Sills \& E. Vesperini, eds.}
\begin{document}

\maketitle

\begin{abstract}
I present the latest results from our group about the multiple
stellar populations in the old Milky Way globular clusters
(GCs) and in the young systems both in the Magellanic Clouds and in
the Milky Way. 
For the ancient GCs in our Galaxy I summarize the chemical
properties of the stellar populations as observed on the chromosome map. Both Type~I and
Type~II GCs are discussed.
For the youngest clusters I will briefly report our latest
spectroscopic analysis on the Large Magellanic Cloud cluster NGC\,1818
and the Galactic open cluster
M\,11, which supports the co-existence of stellar populations with
different rotation rates.
\end{abstract}

\firstsection 

\section{Introduction}

The presence of more than one stellar population in globular clusters
(GCs) is one of the most
fascinating discovery in the field of stellar
populations in the last decade. The definition itself of a GC as the
prototype of a Simple Stellar Population (SSP), definitively shattered
out. One of the direct consequences of the failure of the SSP
assumption is that stellar evolutionary models, which use 
GCs as calibrators, need to be updated. Furthermore, the properties of
stellar populations in these ancient stellar systems provide important
information on the Universe at its earliest phases and on the assembly
of the Milky Way (MW) halo.  
Despite the crucial role played by the presence of more than one
stellar populations in GCs, and the numerous scenarios proposed, at
present none of them is able to properly account for all the
observational constraints (see \cite[Renzini et al.\,2015]{Renzini15}
for a critical discussion).  

Star formation theory assumes that typically stars do not form in
isolation, but in clusters and associations, eventually dissolving.
Star clusters formed along
almost the whole timeline of Universe history, with ancient GCs, like those in the MW,
being among the oldest objects in the Universe. 
Star clusters continue forming and a rich population of young and
intermediate-age objects are observed in the Magellanic Clouds galaxies.

From an observational perspective, the multiple stellar populations
phenomenon is mostly constrained in the ancient MW GCs, that
formed between $\sim$13 and 10~Gyrs ago.
Some hints of different stellar populations are also observed in
intermediate-age objects in the Magellanic Clouds. In this context,
very young clusters are interesting objects because they could
potentially provide a window, in the Local Universe, to observe a poorly-understood phenomenon occurred at
high redshift. The basic question to answer is
if the multiple stellar populations are indeed multiple generations of
stars, and if star clusters could have supported more than one burst
of star formation.

In the following, I will summarize the latest observational results on
multiple stellar populations in the old MW GCs
(Sections~\ref{sec:ChM} and \ref{sec:TyII}), and in some younger
systems both in the Large Magellanic Cloud and in the MW (Section~\ref{sec:young}).

\section{Chemical signatures from the ChMs}\label{sec:ChM}

\begin{figure}[!b]
\begin{center}
 \includegraphics[width=\textwidth]{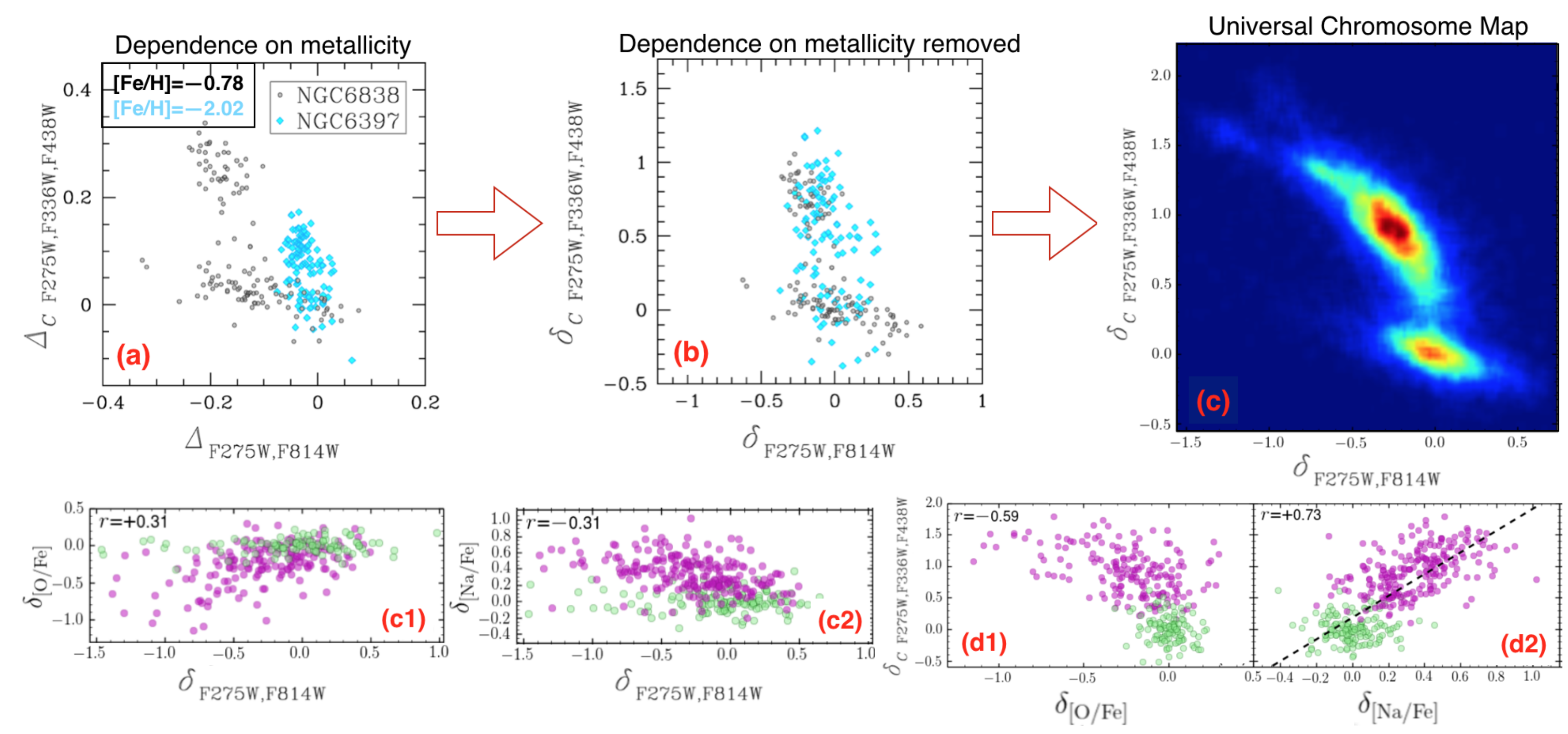} 
 \caption{The universal ChM of GCs (\cite[Marino et al.\,2019]{Mar19}). Panels (a)
   and (b) illustrate the original maps ($\Delta_{\tiny{C~\mathrm{
         F275W,F336W,F438W}}}$
   vs.\,$\Delta_{\tiny{\mathrm{F275W,F814W}}}$), which are 
   metallicity-dependent, and the final maps used to construct the universal
   map ($\delta_{\tiny{C~\mathrm{ F275W,F336W,F438W}}}$
   vs.\,$\delta_{\tiny{\mathrm{F275W,F814W}}}$), where the metallicity
   dependence has been removed. The final 
   universal map is displayed in panel (c). Panels (c1) and (c2) show 
   the abundance ratios
   of O and Na relative to the average abundances of 1G stars vs.
   the $\delta_{\tiny{\mathrm{F275W,F814W}}}$. Panels
   (d1) and (d2) represent 
   $\delta_{\tiny{C~\mathrm{ F275W,F336W,F438W}}}$ as a function of the same abundance ratios. In panels
   (c1)-(c2)-(d1)-(d2) 1G and 2G stars are represented in green
   and magenta, respectively. For each element, we report the
   Spearman's correlation coefficient $r$, and in panel (d2) we plot the
   least-square fit with data.} 
   \label{fig:fig1}
\end{center}
\end{figure}

Central to the characterization of the different stellar populations
in GCs, and for our understanding of this complex phenomenon, is
the knowledge of chemical abundances.
Recently, \cite[Milone et al.\,(2015)]{Mil15} have introduced 
one very effective way of visualizing the complexity of 
multiple stellar populations by means of a diagnostic diagram dubbed
chromosome map (ChM), with such maps having been presented for 58 GCs
(\cite[Milone et al.\,2017]{Mil17}).  
In the ChM plot ($\Delta_{\tiny{C~\mathrm{ F275W,F336W,F438W}}}$
vs.\,$\Delta_{\tiny{\mathrm{F275W,F814W}}}$) 
GCs stars are typically separated into two distinct groups: one
first-population (1G) group and a second-population (2G) one.  
Specifically, 1G stars are located around the origin of the ChM (i.e.,
$\Delta_{\tiny{C~\mathrm{
      F275W,F336W,F438W}}}$=$\Delta_{\tiny{\mathrm{F275W,F814W}}}$=0),
while 2G stars have large $\Delta_{\tiny{C~\mathrm{
      F275W,F336W,F438W}}}$ and low $\Delta_{\tiny{\mathrm{F275W,F814W}}}$. 

One of the most surprising features of the multiple populations
phenomenon, as seen from the ChMs, is its
variety. GCs host different number of stellar populations, with
the extension and morphology of the map changing from one cluster to
another. 
With the aim of trying to find some global pattern among the multiple
populations on the ChM and chemical abundance and investigate possible
common properties in such a variegate zoo, in \cite[Marino et
al.\,(2019)]{Mar19} we attempted a comparison of ChMs of different 
GCs. To this goal we first had to obtain comparable maps, removing
their dependence on metallicity.

The $\Delta_{\tiny{C~\mathrm{ F275W,F336W,F438W}}}$
and $\Delta_{\tiny{\mathrm{F275W,F814W}}}$ width of the ChM
dramatically changes from one cluster to another and mostly correlates
with the cluster metallicity. An example of this dependence is
represented in panel (a) of Fig.~\ref{fig:fig1}.
While both the plotted GCs exhibit a quite simple ChM, 
the ChM extension of the metal-richer cluster (NGC\,6838) is
significantly wider than that of NGC\,6397, which has a much lower
metallicity. After removing this dependence, as explained in
\cite[Marino et al.\,(2019)]{Mar19}, we move to a plane ($\delta_{\tiny{C~\mathrm{ F275W,F336W,F438W}}}$
vs.\,$\delta_{\tiny{\mathrm{F275W,F814W}}}$) where stellar
populations in clusters at different metallicity can be easily
compared (panel (b)).

By combining all the available ChMs in the metal-free plane (excluding
red populations in Type~II GCs which will be discussed in the next
section), we get a ``Universal ChM of multiple stellar populations in
GCs''. Panel (c) displays this universal map as a Hess diagram,
indicative of the density of all the stars. 
The analysis of the main pattern appearing on the universal ChM 
has provided information on the global
properties of the multiple stellar populations phenomenon in GCs.  
On top of the high degree of variety, the universal map displays:
(1) two major over-densities of stars, one on the 1G and the other on
the 2G;
(2) a clear separation between 1G and 2G stars that occurs to
$\Delta_{\tiny{C~\mathrm{ F275W,F336W,F438W}}} \sim$0.25;
(3) the first over-density, the 1G population, appears to occupy a
relatively narrow range in $\Delta_{\tiny{C~\mathrm{
      F275W,F336W,F438W}}}$. The extension in
$\Delta_{\tiny{\mathrm{F275W,F814W}}}$ is larger, but most stars are
located within $-$0.3$< \Delta_{\tiny{\mathrm{F275W,F814W}}} < +$0.3;
(4) the bulk of 2G stars are clustered around
$\Delta_{\tiny{C~\mathrm{ F275W,F336W,F438W}}} \sim$0.9 with a
poorly-populated tail of stars extended towards larger values.  
This suggests that even if the ChMs is variegate, the dominant 2G in
GCs has similar properties. 

The obvious question here is:
how this is related to the chemical composition of stellar
populations? 
The most studied (and distinctive) chemical pattern in GCs is the Na-O
anticorrelation (e.g.\,\cite[Carretta et al.\,2009]{Car09}). 
The universal ChM allowed us to investigate the global variations in
light elements in the overall sample of analyzed GCs.  

In panels (c1) and (c2) of Fig.~\ref{fig:fig1} we plot the abundance
ratios of O and Na relative to the average abundances of 1G stars
($\delta_{\tiny{\mathrm{[O/Fe]}}}$ and $\delta_{\tiny{\mathrm{[Na/Fe]}}}$) as a
function of $\delta_{\tiny{\mathrm{F275W,F814W}}}$. We note that
overall the 2G stars have enhanced Na (higher
$\delta_{\tiny{\mathrm{[Na/Fe]}}}$), and most stars are depleted in O
(lower $\delta_{\tiny{\mathrm{[O/Fe]}}}$). Interestingly, despite the
large range spanned in $\delta_{\tiny{\mathrm{F275W,F814W}}}$, 1G
stars have constant chemical abundance both in O and Na.

Panels (d1) and (d2) represent $\delta_{\tiny{C~\mathrm{
      F275W,F336W,F438W}}}$ as a function of
$\delta_{\tiny{\mathrm{[O/Fe]}}}$ and $\delta_{\tiny{\mathrm{[Na/Fe]}}}$. 
We find a significant correlation with
$\delta_{\tiny{\mathrm{[Na/Fe]}}}$. 
An anticorrelations is present between $\delta_{\tiny{C~\mathrm{
      F275W,F336W,F438W}}}$ and $\delta_{\tiny{\mathrm{[O/Fe]}}}$.

While I refer to \cite[Marino et al.\,(2019)]{Mar19} for a detailed
chemical characterization of the GC ChMs, here I emphasize that the presented
analysis demonstrates the power of
the ChM
to disentangle between stars with different
chemical abundances. Hence, individual ChMs and the
universal ChM are, at present, the best tools available to explore the
multiple stellar population phenomenon and the nature of the different
populations hosted in GCs.

\section{Signatures of heavy element variations}\label{sec:TyII}

\begin{figure}[!b]
\begin{center}
 \includegraphics[width=\textwidth]{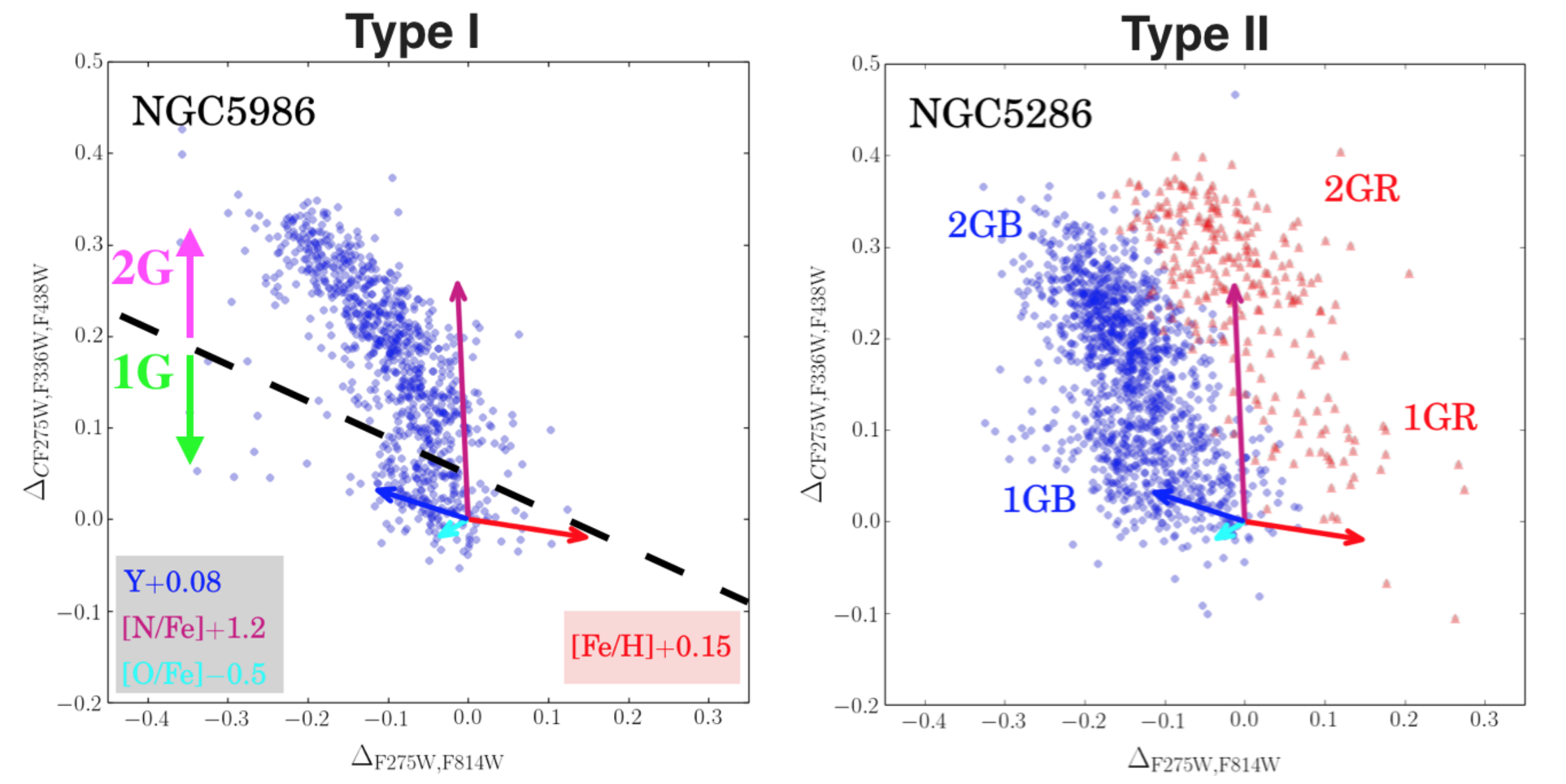} 
 \caption{ChMs of the Type~I GC NGC\,5986 and the
   Type~II GC NGC\,5286. The arrows indicate the effect of
   changing He, N, O, and Fe by the quoted quantities. The diagram of Type~II GCs
   shows additional maps (1GR and 2GR) on the red side of
   the main map (1GB and 2GB). The dashed line on the left panel
   separates 1G from 2G stars.}
   \label{fig:TyI-TyII}
\end{center}
\end{figure}

\begin{figure}[!b]
\begin{center}
 \includegraphics[width=\textwidth]{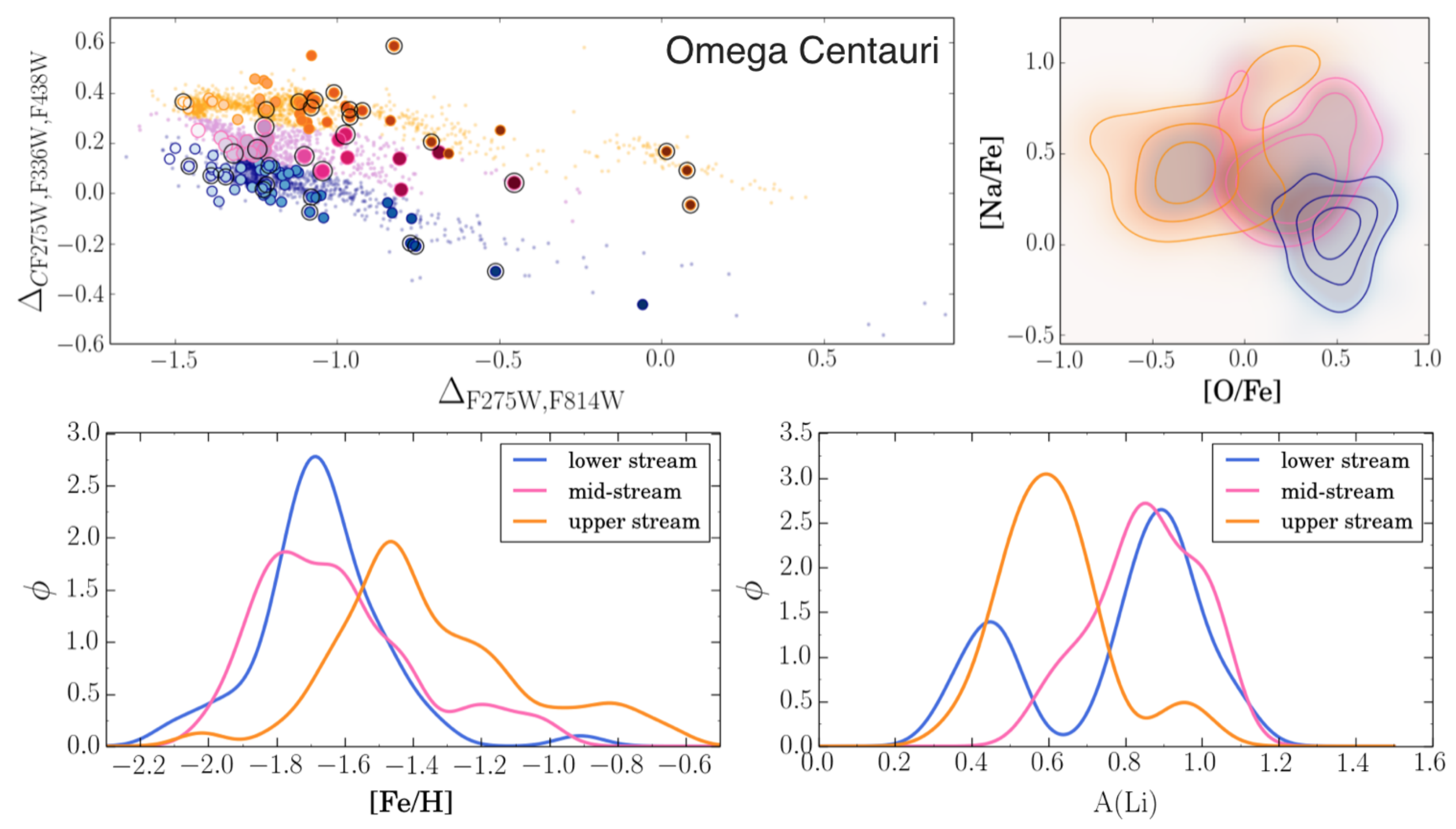} 
 \caption{{\it Top-left panel}: ChM of $\omega$~Centauri. 
   The lower, middle, and upper streams have been
   coloured in light blue, pink and orange, respectively. 
   Larger dots are stars with abundances from \cite[Johnson \& Pilachowski\,(2010)]{J&P10} and
   \cite[Marino et al.\,(2011)]{Mar11} 
   (black-circled points). To each star with abundances a gradient in
   colour has been assigned proportional to its $\Delta_{\tiny{\mathrm{F275W,F814W}}}$. 
   {\it Top-right panel}: density levels on the Na-O plane for stars belonging to each stream. 
   {\it Lower panels}:
   Gaussian kernel density distributions of [Fe/H] and A(Li)
   for the streams by using the \cite[Johnson \& Pilachowski\,(2010)]{J&P10},
   \cite[Marino et al.\,(2011)]{Mar11} and \cite[Mucciarelli et al.\,(2018)]{Mucc18} samples.}  
   \label{fig:OmegaCen}
\end{center}
\end{figure}

The variety of the multiple populations phenomenon has prompted
efforts to classify GCs in different classes, whose definition has
changed from time to time, depending on the new features being
discovered. 
\cite[Milone et al.\,(2017)]{Mil17} subdivided the analyzed ChMs in
two main groups: 
{\it (i)} Type~I GCs having a single ChM pattern, with a 1G and two or
more 2G populations, and {\it (ii)} Type~II GCs displaying multiple ChMs,
with minor populations located on red additional ChMs. The ``Type~II
phenomenon'' occurs in $\sim$17\% of the analyzed clusters, and
suggests that these objects have experienced a much more complex star
formation history. 

In Fig.~\ref{fig:TyI-TyII} I show the ChMs of a typical Type~I GC,
NGC\,5986, and a typical Type~II one, NGC\,5286. Other examples of
Type~II GCs are M\,22, M\,2, NGC\,6934, and $\omega$~Centauri. The
analysis of the effect of variations in chemical elements on the map,
clearly shows that the increase in the
$\Delta_{\tiny{C~\mathrm{F275W,F336W,F438W}}}$ index is primarily due
to nitrogen enhancements, while
helium increases with decreasing $\Delta_{\tiny{\mathrm{F275W,F814W}}}$
(\cite[Milone et al.\,2015]{Mil15}).

The $\Delta_{\tiny{\mathrm{F275W,F814W}}}$ index is sensitive to the
overall metallicity, suggesting that
stars on the red sequences of the map are enhanced in [Fe/H]. Indeed,
all the GCs with known variations in iron
belong to this (photometrically defined) class of objects
(e.g.\,\cite[Marino et al.\,2018]{Mar18}). In most of these clusters,
internal variations in $slow$-neutron capture elements correlate
with the enhancement in metals
(e.g.\,\cite[Marino et al.\,2015]{Mar15}). 
The reason why this different chemical behavior occurs remains to be established, but,
intriguingly, might be linked to a different origin with respect to
the more common GCs in the Galaxy (see discussion in \cite[Marino et al.\,2019]{Mar19}). 

Omega Centauri, with its well-documented large variations in metallicity, is
the most extreme example of a Type~II GC. 
We have explored in details the ChM of this GC, whose wide ranges
in chemical abundances make it an ideal laboratory to
test the role of each element in shaping the GC ChMs.  

In Fig.~\ref{fig:OmegaCen} three main streams of stars are evident at
different levels of $\Delta_{\tiny{C~\mathrm{F275W,F336W,F438W}}}$. 
The upper stream stars mostly occupy the section of higher Na and
lower O quadrant of the O-Na plane with [O/Fe]$<$0.0~dex. 
Mid-stream stars mostly distribute on a intermediate location in the O-Na
plane, while lower stream stars have lower Na and higher O. 
As a general rule, stellar populations with increasing Fe
populate redder and redder regions of the map.
The Fe enrichment seems generally de-coupled from the light-elements
processing. The lower and mid-streams are
peaked at similar Fe ([Fe/H]$\sim -$1.7~dex), with the
mid-stream having a minor over-density of stars at higher Fe.
The upper stream is the most distinct: it is peaked on higher
metallicity and lower Li abundances.
As the upper-stream red-RGB stars have a more extreme enrichment in Na and
Al, and depletion in Li and O, these stars are likely the
most-enhanced in helium. 

This discussion highlights the efficiency of the
ChMs in separating stars with different chemical content
in heavy elements. This diagnostic tool can be exploited to identify
GCs with variations in the overall metallicity.
A chemical abundance exploration of the ChM of $\omega$~Centauri suggests
that the most important Fe-enrichment occurred when the intra-cluster
medium was already extremely enriched in He and in H-burning products.

\begin{figure}[!htp]
\begin{center}
 \includegraphics[width=0.95\textwidth]{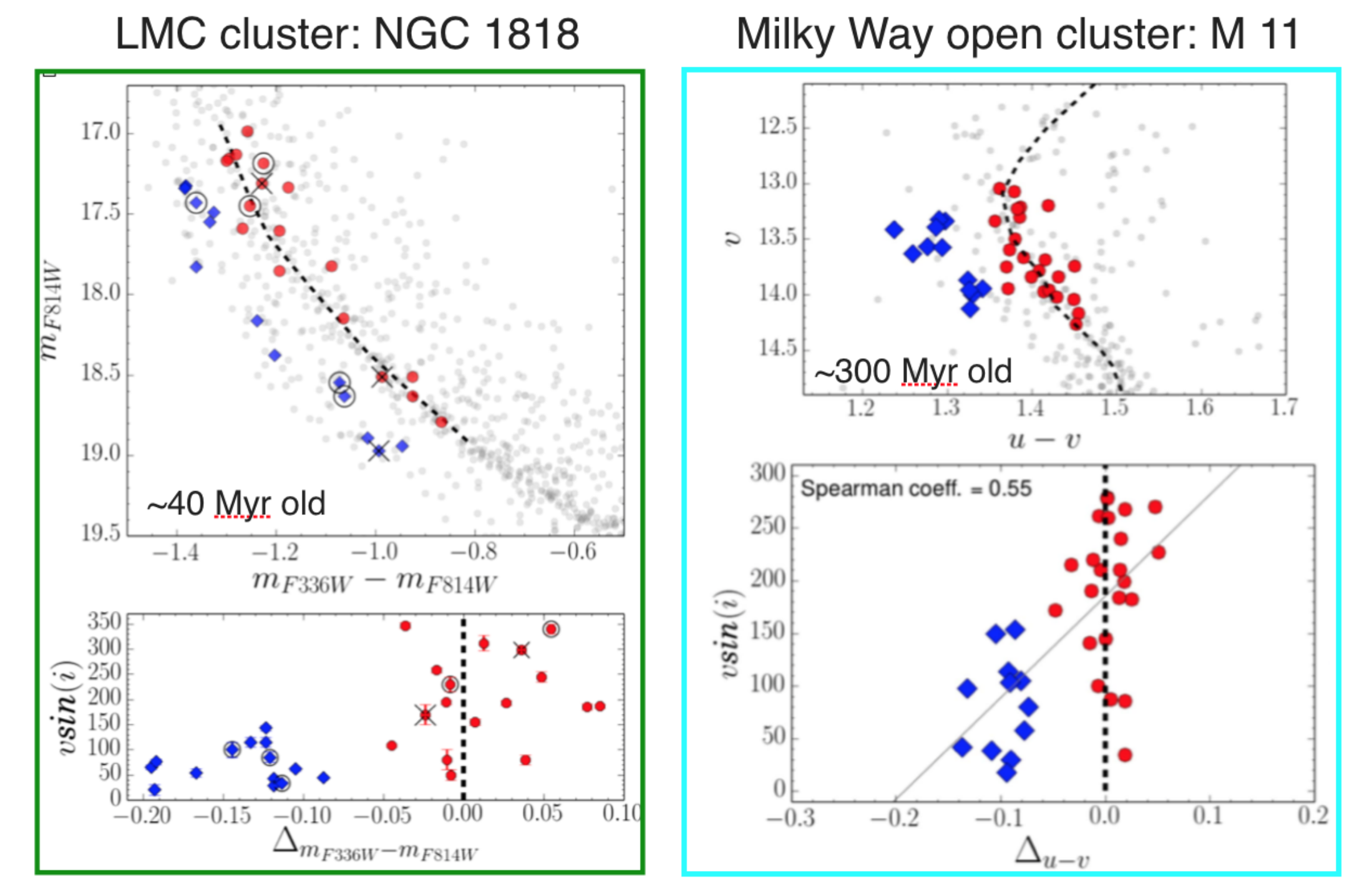} 
 \caption{{\it Left panels}: $m_{\mathrm F814W}$-$(m_{\mathrm F336W} -
   m_{\mathrm F814W})$ CMD of the LMC cluster NGC\,1818 zoomed-in on
   the split MS. The black dashed line is a fiducial for the rMS
   (upper panel). The lower panel shows $v$sin$i$ as a function of the
   difference between the $m_{\mathrm F336W} - m_{\mathrm F814W}$
   color of each MS star and the color of the fiducial
   ($\Delta_{m_{\mathrm F336W} - m_{\mathrm F814W}}$). 
   The circle and cross symbols indicate SB1 candidates and stars with H$\alpha$
   core and/or wings emissions, respectively. {\it Right panels}:
   $v$-$(u-v)$ CMD of the open cluster M\,11 zoomed-in on 
   the split MS. The black dashed line is a
   fiducial for the rMS. $v$sin$i$ as a function of the
   difference between the $u - v$
   color of each MS star and the color of the fiducial
   ($\Delta_{u-v}$) is shown in the lower panel. In all the plots, rMS
   and bMS stars are represented
   with red filled circles and blue diamonds, respectively.}
   \label{fig:young}
\end{center}
\end{figure}

\section{Young clusters in the Magellanic Clouds and the Milky Way}\label{sec:young}

It seems that multiple stellar populations are not a peculiarity of
the old Galactic GCs. Indeed, 
split main sequences (MSs) and extended MS turn-offs (eMSTOs)
have been discovered along the 
color-magnitude diagrams (CMDs) of young (age $\sim$50-500~Myr)
clusters in both Magellanic 
Clouds (e.g.\,\cite[Milone et al.\,2016]{Mil16}).
Despite a huge effort has been undertaken to understand these
observations, the physical mechanism responsible for the eMSTO is
still obscure. 

The eMSTO has  been originally interpreted as due to an age spread of
50-500~Myrs; in such a case the clusters with the eMSTO would be the younger counterparts of the
old GCs with multiple populations (\cite[Mackey \& Broby Nielsen\,2007]{M&BN}). Another interpretation is
that the eMSTO is an effect of stellar rotation (e.g.\,\cite[Bastian
\& de Mink\,2009]{B&dM}).

We have reported the first direct spectroscopic measurements of
projected rotational velocities ($v$sin$i$) for the double MS, and
eMSTO in a Large Magellanic Cloud (LMC) cluster, namely NGC\,1818
(\cite[Marino et al.\,2018b]{Mar18b}). Figure~\ref{fig:young} shows
that the blue-MS (bMS) is populated by slowly rotating stars, while
the red-MS (rMS) is composed of fast rotators, with rotation close to
the breaking speed (left panel of Fig.~\ref{fig:young}). 
Similar multiple sequences are observed in the Milky Way open clusters
at similar ages (\cite[Marino et al.\,2018c]{Mar18c}; \cite[Cordoni et
al.\,2018]{Cordoni18}). Also in the Milky Way, this phenomenon
is associated to stellar populations with different rotation (right
panel of Fig.~\ref{fig:young}; \cite[Marino et al.\,2018c]{Mar18c}). 
These observations suggest that multiple sequences in young clusters are
populated by stars with different rotation, and we might not need an
age spread to reproduce the observed CMDs.

\section*{Acknowledgments}
\small
I warmly thank all my collaborators.
This work has received funding from the European Union's Horizon 2020
research and innovation programme under the Marie Sk{\l}odowska-Curie
(Grant Agreement No. 797100).

\end{document}